\documentstyle[11pt]{article}
\def\mgluon{\langle 0 | \frac{\alpha_{s}}{4 \pi } G \tilde{G}
 | \pi^{0} \rangle }
\def\lpi{\langle\pi^{0} |}
\def\lo{\langle 0 |}
\def\ro{ | 0 \rangle }
\def\rpi{|\pi^{0} \rangle }
\def\lpi{ \langle \pi^{0} | }
\def\gmmu{\gamma _{\mu}}
\def\atop{ \frac{ \alpha_{s}}{4 \pi} G \tilde{G} }
\def\gmf{\gamma _{5}}
\def\ratio { \frac{m_{d}-m_{u}}{m_{d}+m_{u}} }
\def\ll{\langle }
\def\rl{ \rangle }
\def\pb{ | \pi_{B} \rl}
\def\etb{ | \eta_{B} \rl}
\def\et'b{ | \eta'_{B} \rl}
\newcommand{\beq}{\begin{equation}}
\newcommand{\eeq}{\end{equation}}
\newcommand{\bea}{\begin{eqnarray}}
\newcommand{\eea}{\end{eqnarray}}
\newcommand{\preprint}[1]{\begin{table}[t]
           \begin{flushright}
           \begin{large}{#1}\end{large}
           \end{flushright}
           \end{table}}
\begin{document}
%
\preprint{TAUP-2127-93 \\ December 1993}
\title{Veneziano Ghost Versus Isospin Breaking }

\author{\\ Igor Halperin\thanks{Bitnet:IGOR@TAUNIVM.} \\ \ }
\date{ School of Physics and Astronomy \\ Raymond and Beverly Sackler
Faculty of Exact Sciences \\
Tel-Aviv University, Tel-Aviv 69978, Israel}
\maketitle
\begin{abstract}
It is argued that an account for the Veneziano ghost pole, appearing
in resolving the U(1) problem, is necessary for understanding an isospin
violation in the $ \pi - \eta - \eta' $ system. By virtue of a
perturbative expansion around the $ SU(2)_{V} $ ( $ m_{u} = m_{d} $ )
symmetric Veneziano solution, we find that the ghost considerably
suppresses isospin breaking gluon and s-quark matrix elements.
We speculate further on a few cases where the proposed mechanism can
play an essential role.
We discuss the isospin violation in meson-nucleon couplings
and its relevance to the problem of charge asymmetric nuclear forces
and possible breaking of the  Bjorken sum rule.
It is shown that the ghost pole could yield the isospin violation
of order 2 \% for the $ \pi N $ couplings and 20 \% for the
 Bjorken sum rule.
\end{abstract}
\section{Introduction}
The isospin symmetry is known to be a good approximation to the real
world. The common wisdom claims that one can set $ m_{u} \simeq m_{d} $
not because of their closeness ( $ m_{d} - m_{u} \sim O(m_{u},m_{d}) $ )
but rather due to the fact that they are both small at the scale of
hadronic masses. For the case of pseudo-goldstone bosons, this scale is
set, however, by the quark masses themselves, that makes their physics
exceptional in the following sense. Since $ \frac{m_{d}-m_{u}}{m_{d}+
m_{u}} \sim O(1) $ , one could expect large isospin violations which are
not usually observed experimentally. On the other hand, there exist
phenomena (e.g. heavy quarkonia decays, isospin symmetry breaking in
nuclea, etc.) where these effects may be significant.

As it has been suggested by Gross, Treiman and Wilczek (GTW)
in their seminal paper [1], the
reason for small isospin breaking lies in the existence of the axial
anomaly. By its virtue, the $\eta '$ gains a large mass and effectively
decouples from mixing in the $\pi^{0}-\eta-\eta' $ system. In this case,
 an isospin violation is only $ O( \frac{m_{d}-m_{u}}{m_{s}}) $ due to
$\pi^{0}-\eta $ mixing (under assumption on the flavor independence for
condensates). However, the $\eta'$ has a mass comparable to that of the
$\eta$ and sizable mixing with the $\eta$ . Thus, one can conclude that
an account for the mass generation for the $\eta'$ (i.e. a solution of
the celebrated U(1) problem) is necessary for quantitative studying the
isospin violating effects. ( In all likelihood, one can neglect
electromagnetic interactions for isospin breaking since $(\frac{m_{d}-
m_{u}}{m_{s}})/ \frac{ \alpha}{ \pi} \sim 10 $ ).

 In this paper, we apply the Veneziano scheme of resolving U(1) problem
[2,3] , combined with a perturbation theory over $ \frac{m_{d}-m_{u}}{
m_{d}+m_{u}} $
,to the problem of mixing $ \pi^{0}-\eta-\eta'$ . In Sect.2 we discuss
problems with calculating the pion matrix element of the topological
density which plays an important role in the heavy quarkonia decays, see
e.g. [5,6] . The Veneziano construction in the $ SU(2)_{V} $ ( $m_{u}=
m_{d} $ ) limit is reviewed in Sect.3 , the presentation being close to
that of Ref.[3]. In Sect.4 we develop a perturbation theory around this
limit and calculate first order corrections to the wave functions. This
allows us to re-estimate the gluon matrix element discussed in Sect.2.
( It should be noted that the only known to the author anomaly-based
discussion of the role of mixing in the gluon matrix elements [4] has
rested on adding "by hands" the mass terms into the effective chiral
Lagrangian and seems to contradict in both modulo and the signs
to the estimates made in [1,3] ). The proposed technique is applied
to a calculation of s-quark matrix elements in Sect.5 . We proceed
further in Sect.6 to discussing the isospin asymmetry in the $ \pi N $
 interaction resulting from the presented mechanism of the mixing.
It should be stressed that we only concern with the ghost contribution
into the studied phenomena. Nonanomalous mechanism are left beyond the
scope of this paper and deserve separate studying. The effects
are found to be sizable, of order 2 \% , while their sign turns
out to be opposite to what is expected for a solution of the so-called
Nolen-Schiffer anomaly known in nuclear physics [15,16,23] . A relevance
to the proton spin problem and possible violation of the Bjorken sum rule
in QCD is discussed in Sect.7.
Sect.8 contains some comments and conclusions.

\section{ Matrix element $ \lo \alpha_{s} G \tilde{G} | \pi^{0} \rangle }

There are a few reasons to start our presentation with this
matrix element. The first
one, it is proportional to the isospin breaking parameter $ \frac{m_{d}
-m_{u}}{m_{d}+m_{u}} $ and related to the experimental heavy quarkonia
decay ratios like $ \frac{ \Gamma ( \Psi' \rightarrow \Psi + \pi^{0})}
{ \Gamma ( \Psi' \rightarrow \Psi + \eta )} $ [5,6]. The second one, and
more important for us, it helps to explain the necessity for an account
for mixing for its correct determination. Using the soft pion technique
and expanding to the first order in $ L^{ \Delta I=1} = - \frac{m_{u}-m_{
d}}{2} ( \bar{u}u - \bar{d}d ) $ , one can write the following chain of
equations
\bea
\mgluon & = & i \int dy e^{iqy} ( \Box_{y}+ m^{2}_{\pi}) \lo T \{ \phi
_{\pi^{0}}(y) \frac{ \alpha_{s}}{4 \pi} G \tilde{G} (0) \} | 0 \rangle
  \\  \nonumber
        & = & _{q_{\mu} \rightarrow 0}  \frac{i}{\sqrt{2} f_{\pi}}
 \int dy
\lo T \{ \partial _{\mu} J^{5}_{ \mu}(y) \atop (0) \} \ro  \\ \nonumber
        & = & \frac{m_{u}-m_{d}}{2 \sqrt{2} f_{\pi}} \int dz dy \lo
T \{ \partial^{(y)}_{\mu} J^{5}_{\mu}(y) (\bar{u}u - \bar{d} d )(z)
\atop (0) \} \ro  \\
        & = & \frac{m_{u}-m_{d}}{ \sqrt{2} f_{\pi}} \int dz \lo T \{
  ( \bar{u}
\gmf u + \bar{d} \gmf d)(z) \atop (0) \} \ro  \nonumber \; ,
\eea
where on the last step we have used the Ward identity with the
canonical commutator $ \delta (y_{0}-z_{0}) [J^{5}_{0}(y)\;,\;
(\bar{u}u- \bar{d}d)(z) ] = -2 \delta (y_{0}-z_{0}) \delta ^{3}
( \vec{y}- \vec{z} ) [ \bar{u} \gmf u + \bar{d} \gmf d ] (z) \;,
\; J^{5}_{ \mu} = \bar{u} \gamma _{\mu} \gmf u - \bar{d} \gamma
_{ \mu} \gmf d $  and the PCAC formula  $ \phi_{\pi ^{0}}(x)=
\frac{1}{ \sqrt{2} f_{\pi} m^{2}_{\pi}} \partial _{\mu} J^{5}_{\mu}
(x)  $ .   The last correlator can now be extracted from the low
energy theorem  [7] that yields the result ( $ f_{ \pi} \simeq 133
 \; Mev $ )
\beq
\mgluon \; =\; \frac{1}{ \sqrt{2}} f_{\pi} \frac{ m_{d}-m_{u}}
{m_{d}+m_{u}} m^{2}_{\pi} ( 1\;+\; \zeta )  \; ,
\eeq
( here $ \zeta $ is a correction factor which we will discuss later on)
that coincides with the answer [1] obtained for this matrix element
by a different method. However, the above derivation can ( and, in fact,
must ) be seriously criticized. First, a non-zero mass difference
$ m_{u}- m_{d} $ induces mixing of the $ \pi_{0} $ with $ \eta\;,\;
\eta' $ , so that the second equality  in (1) seems rather suspicious.
 There one may expect an error of order
 50 \% since a small mixing angle  $ O( \frac{ m_{d}-m_{u}}{m_{s}})
$ can be well compensated by a large value of $ \lo \atop | \eta
\rangle \sim f_{\pi} m^{2}_{\eta} $ [8] with $ \frac{m^{2}_{ \eta}}
{m^{2}_{\pi}} \sim \frac{m_{s}}{ m_{u}+ m_{d}} $ . Second, one may
wonder whether the account for $ L^{ \Delta I=1} $ generates new
interactions involving the Veneziano ghost [2] potentially revealing
itself in the correlator (1).

Let us address now the original derivation of GTW [1]. They have used
the Sutherland theorem [9] for the SU(2) singlet current
$ j_{\mu} \;=\; \bar{u} \gamma _{\mu} \gmf u \;+\; \bar{d} \gamma _{\mu}
 \gmf d $ , stating that
\beq
\lim_{ q_{\mu}  \rightarrow 0} \; q_{\mu} \int dx e^{iqx} \langle
\gamma (k_{1}) \gamma (k_{2}) | j_{\mu} \ro \;=\; 0
\eeq
After saturating this correlator by the pion intermediate state and
accounting for the anomaly's contribution, they have arrived at the
equation
\beq
\mgluon \;=\;  - \lo m_{u} \bar{u}i \gmf u + m_{d} \bar{d} i \gmf
d \rpi
\eeq
To discuss possible corrections to (4) , consider the anomaly equation
sandwiched in between the vacuum and one pion states
\beq
- i q_{\mu} \lo j_{\mu} \rpi = A f_{\pi} m^{2}_{\pi}  =
 2( \lo m_{u} \bar{u} i \gmf u + m_{d} \bar{d} i \gmf d \rpi
 + \mgluon  ) \;,
\eeq
where we have defined $ \lo \bar{u} \gamma _{\mu} \gmf u + \bar{d}
\gamma_{\mu} \gmf d \rpi = i f_{\pi} A q_{\mu} $ . The unknown parameter
A has to be of the form $ - \frac{1}{\sqrt{2}} \frac{ m_{u}-m_{d}}{m_{u}+
m_{d}} \zeta $  where $ \zeta $  is some dimensionless
function of the mass
parameters. Along with reasoning of GTW, one has to expect $ \zeta \sim
O( \frac{m_{q}}{M_{st}}) $ that can yet bring in corrections of order
30 \% due to $ \frac{m_{s}}{M_{st}} $  [1,5] ( $ M_{st} $ stands for a
scale of the strong interaction ). However, such corrections apparently
do not show up in eq. (4) . This seeming paradox is traced back to the
fact that eq. (4) stands for the {\it off-shell} matrix elements. While
the mixing effects on the matrix element of the isoscalar density in the
r.h.s. of eq. (4) are small {\it on} the mass shell, this property can be
lost for the off-shell case $ q_{\mu} \rightarrow 0 $ , thus bringing
back the $ \zeta $ factor into the  on-shell version of eq.(4). The
two above derivations yield the same answer as long as they are obtained
within the same approximation , i.e. they both neglect mixing as well as
the on-shellness. As will be argued in Sect.4, there is {\it another}
class of corrections which is not taken into account in (2),(4). It is
the Veneziano ghost [2] that brings a leading large ( of order 40 \% )
and negative correction to the GTW formula (2).

\section{Ghost pole and the U(1) problem}

Let us start with introducing the following currents
\beq
J^{1}_{\mu 5}= \frac{1}{\sqrt{2}} ( \bar{u} \gmmu \gmf u +
\bar{d} \gmmu \gmf d ) \:, \: J^{2}_{\mu 5 }= \bar{s} \gmmu \gmf s \:,
\: J^{3}_{\mu 5}= \frac{1}{\sqrt{2}} ( \bar{u} \gmmu \gmf u - \bar{d}
\gmmu \gmf d )
\eeq
with
\bea
\partial_{\mu} J^{1}_{\mu 5} & = & i \sqrt{2} ( m_{u} \bar{u} \gmf u
+ m_{d} \bar{d} \gmf d ) + \sqrt{2} \atop  \equiv P_{1} + 2 \sqrt{2}
Q          \nonumber  \\
\partial_{\mu} J^{2}_{\mu 5} & = & 2 i m_{s} \bar{s} \gmf s + 2 Q
\equiv P_{2} + 2 Q   \\
\partial_{\mu} J^{3}_{\mu 5} & = & i \sqrt{2} ( m_{u} \bar{u} \gmf
u - m_{d} \bar{d} \gmf d ) \equiv P_{3}    \nonumber
\eea
This choice of the currents is motivated by the fact that they
correspond to the mass eigenstates
\beq
\pi_{1} \sim \frac{ \bar{u} \gmf u + \bar{d} \gmf d }{ \sqrt{2}} \;,
\; \pi_{2} \sim \bar{s} \gmf s \; ,\; \pi_{3} \sim \frac{ \bar{u} \gmf u
- \bar{d} \gmf d }{ \sqrt{2}}
\eeq
with  ( $ f_{1} \; = \; f_{\pi} \;,\; m_{q}=\frac{m_{u}+m_{d}}{2} \; ,
\; \langle \bar{u} u \rangle = \langle \bar{d} d \rangle =
\langle \bar{q} q  \rangle  $   )
\bea
m^{2}_{3} & = & m^{2}_{1} = - \frac{1}{f^{2}_{1}} 4 m_{q}
\langle \bar{q} q \rangle + O ( m^{2}_{q}) \simeq 0.02 \; Gev^{2} \\
m^{2}_{2} & = & -\frac{1}{f^{2}_{2}} 4 m_{s} \langle \bar{s} s
\rangle + O ( m^{2}_{s} )  \simeq m^{2}_{K^{0}} + m^{2}_{K^{-}}
- m^{2}_{\pi^{0}} \simeq 0.47 \; Gev^{2}      \nonumber
\eea
in the limit where the both quark mass difference and anomaly are
neglected  ( Clearly, the formulae (9),(10) do not respect the real
world with $ m_{\eta} \simeq 549 \; Mev $  and  $ m_{\eta'} \simeq
958 \; Mev $ ). In the naive chiral limit   $ m_{q}\;=\;0 $
the formula (8) represents the composite goldstone fields related
to the spontaneous breaking of the chiral U(3) invariance  $ \langle
\bar{q} q \rangle  \neq 0 $  .

Making use of the standard technique [10], one can obtain the following
set of the Ward identities (WI) ( hereafter we denote $ \langle A \: B
\rangle _{q} \:= \: i \int dx e^{iqx}  \langle T \{ A(x) B(0) \} \rangle
 \;,\; \langle A \: B \rangle \: = \: \lim_{ q \rightarrow 0} \langle
 A \: B \rangle_{q}  $  )
\bea
\ll P_{3} P_{3} \rl \; + \; 2 m_{u} \ll \bar{u} u \rl  \; + \;
 2 m_{d} \ll \bar{d} d \rl \;=\; 0  \\
\ll P_{3} P_{1} \rl \; + \; 2 m_{u} \ll \bar{u} u \rl \; - \;
2 m_{d} \ll \bar{d} d \rl \; = \; 0    \\
\ll P_{2} P_{3} \rl \; = \; 0         \\
\ll Q P_{3} \rl \; = \; 0        \\
\ll P_{1} P_{1} \rl  \: +  \: 2 \sqrt{2} \ll Q P_{1} \rl   \: + \:
2 m_{u} \ll \bar{u} u \rl \: + \: 2 m_{d} \ll \bar{d} d \rl \: =
\: 0                          \\
\ll P_{1} P_{2} \rl \; + \; 2 \sqrt{2} \ll Q P_{2} \rl \; = \; 0  \\
\ll Q P_{1} \rl \; + \; 2 \sqrt{2} \ll Q Q  \rl \; = \; 0    \\
\ll P_{2} P_{2} \rl \; + \; 2 \ll Q P_{2} \rl \; + \;
4 m_{s} \ll \bar{s} s \rl \; = \; 0        \\
\ll Q P_{2} \rl \; + \; 2 \ll Q Q \rl \; = \; 0
\eea
We immediately note that in the $ SU(2)_{V} $ limit  $ m_{u} =
m_{d} $  the $ \pi_{3} $ state decouples from the anomalous WI (14-18),
thus simplifying enormously their analysis. Therefore we will first
concentrate on this limit and postpone a discussion of isospin breaking
corrections until Sect.4 .

As one can see from (14),(16) , the correlator of the topological
density  Q  at zero momentum  $ \ll Q Q \rl  $ must have a non-zero
value in order to repair the wrong mass formulae (9). It can be shown
from the analysis of the commutation relations that the correlator
$ \ll Q Q \rl $ entering the WI (14)-(18) is to be recognized as
defined by virtue of the Wick T-product [3]. Moreover, the following
relation holds
\beq
\ll Q Q \rl ^{W}_{q} \: = \: \ll Q Q \rl ^{D}_{q} - \ll \frac{ \alpha_{s}
}{ 4 \pi} G^{2} \rl \: = \: (-iq_{ \mu} ) ( - iq_{ \nu} )
\ll K_{ \mu} K_{ \nu} \rl ^{D}_{q}
\eeq
( The symbols D , W stand for the Dyson and Wick T-products ). Here
\beq
K_{\mu} = \frac{ \alpha_{s}}{ 4 \pi} \varepsilon_{ \mu \alpha \beta
\gamma } A^{a}_{ \alpha} (  \partial_{\beta} A^{a}_{\gamma} +
\frac{1}{3} g f^{abc} A^{b}_{\beta} A^{c}_{ \gamma} ) \: , \:
Q = \partial_{\mu} K_{\mu}
\eeq
is the gauge-dependent gluon current. Thus, the presence of a pole
as $ q^{2} \rightarrow 0 $ in the correlator $ \ll K_{\mu} K_{\nu}
\rl $  is necessary for a solution of the U(1) problem [2,3] :
\beq
\ll K_{\mu} K_{\nu} \rl ^{D}_{q} \; = \; const \: \frac{ g_{\mu \nu }}
{ q^{2} }
\eeq
( It is worth recalling that only the Dyson T-product admits a
intermediate states representation ). The residue in this pole can
be argued to vanish in the $ m_{q} \rightarrow 0 $  limit. As has
been shown in [3] , the existence of such ghost pole is deeply
motivated :  it is the consequence of a periodic dependence of the
potential energy in QCD on the collective variable  $ X = \int
d^{3} \vec{x} K_{0} ( \vec{x} , t) $ . Thus, its appearance is a
 purely nonperturbative phenomenon.

After this physical idea is introduced, mathematics becomes rather
simple. One defines [2,3] the ghost propagator
\beq
\ll a_{\mu} a_{\nu} \rl_{0} \; = \; - \frac{q_{\mu} q_{\nu}}{q^{4}}
\eeq
with
\beq
\lo K_{\mu} | a^{p} \rl \; = \; \lambda^{2} \varepsilon^{p}_{\mu}
\; , \; \lo  Q  | a_{\mu} \rl \; = \; -i q_{\mu} \lambda^{2}
\eeq
( $ \varepsilon ^{p}_{\mu} $  stands for the polarization vector ) , so
that
\beq
 \ll Q Q \rl ^{W} \; = \; - \lambda^{4} \; \neq \; 0
\eeq
One introduces further the bare propagators for the $ \pi_{1,2} $
states
\beq
\ll \pi_{i} \pi_{j} \rl_{0} \; = \; \delta_{ij} \frac{1}{ m^{2}_{i}
- q^{2}}
\eeq
( i.e. those obtained when gluon intermediate states are excluded )
and the point-like transition amplitudes
\beq
\ll a_{ \nu}  | \pi_{i} \rl \; = \; - i q \mu _{i}
\eeq
Then the exact propagators can be found from (22-26) by solving the
system of the Dyson equations ( no sum over i )
\bea
\ll \pi_{i} \pi_{i} \rl & = & \ll \pi_{i} \pi_{i} \rl_{0}
(1+ \ll \pi_{i} | a_{ \mu} \rl \ll a_{ \mu} a_{ \nu  } \rl
\ll a_{ \nu} | \pi_{i} \rl \ll \pi_{i} \pi_{i} \rl_{0}) \nonumber \\
\ll \pi_{i} \pi_{j} \rl & = & \ll \pi_{i} \pi_{i} \rl_{0}
\ll \pi_{i} | a_{ \mu} \rl \ll a_{ \mu } a_{ \nu} \rl \ll a_ { \nu}
 | \pi_{j} \rl \ll \pi_{j} \pi_{j} \rl_{0}   \\
\ll a_{ \mu} a_{ \nu} \rl & = & \ll a_{ \mu} a_{ \nu} \rl_{0} +
\sum_{i} \ll a_{ \mu} a_{ \rho} \rl_{0} \ll a_{ \rho} | \pi_{i} \rl
\ll \pi_{i} \pi_{i} \rl \ll \pi_{i} | a_{ \xi} \rl \ll a_{ \xi} a_{ \nu
} \rl_{0}     \nonumber  \\
\ll a_{ \mu} \pi_{i} \rl & = & \ll a_{ \mu} a_{ \nu} \rl \ll a_{ \nu} |
\pi_{i} \rl \ll \pi_{i} \pi_{i} \rl_{0}   \nonumber
\eea
Then, all the masses, wave functions and matrix elements can be
calculated from successive saturating the gauge invariant correlators
by the contributions of $ \pi_{1,2} , a_{ \mu} $  and physical $ \eta ,
\eta' $ states, with the parameters $ \mu_{i} , \lambda^{2} $ being
defined with the help of the WI (14-18) ( plus e.g. the experimental
value for $ m^{2}_{ \eta} + m^{2}_{ \eta'} $  [2,3]). We just cite the
final answers [3]
\bea
m^{2}_{ \eta , \eta'} = \frac{1}{2} \{ ( m^{2}_{1} + m^{2}_{2} +
\mu^{2}_{1} + \mu^{2}_{2} )  \pm  \sqrt{ ( m^{2}_{1} + \mu^{2}_{1}
- m^{2}_{2} - \mu^{2}_{2} )^{2}  + 4 \mu^{2}_{1} \mu^{2}_{2} } \:
\}      \nonumber  \\
\frac{f_{1} \mu_{1}}{ \sqrt{2}}=f_{2} \mu_{2}= 2 \lambda^{2} \: , \:
 \mu^{2}_{1} \simeq 0.57 \: Gev^{2}  \: , \:  \mu^{2}_{2} \simeq
0.16 \: Gev^{2}    \\
m^{2}_{ \eta} \simeq 0.307 ( exp. 0.301) \: Gev^{2} \; , \;
m^{2}_{ \eta'} \simeq 0.912 ( exp. 0.917) \: Gev^{2}   \nonumber
\eea
\bea
| \eta' \rl & = &  \cos \theta | 1 \rl + \sin \theta | 8 \rl  \\
| \eta \rl  & = & - \sin \theta | 1 \rl + \cos \theta | 8 \rl
\; \: ( \theta \simeq - 10 ^{0} )   \nonumber
\eea
and
\bea
\lo Q | \eta \rl = \lambda^{2} \sqrt{ (m^{2}_{2} -m^{2}_{ \eta} )
( m^{2}_{ \eta} - m^{2}_{1} ) / ( m^{2}_{ \eta'} - m^{2}_{ \eta})}
\simeq 0.010 \;Gev^{3}  \\
\lo Q | \eta' \rl = \lambda^{2} \sqrt{ (m^{2}_{ \eta'} - m^{2}_{2})
( m^{2}_{ \eta'} - m^{2}_{1} ) / ( m^{2}_{ \eta'} - m^{2}_{ \eta})}
\simeq 0.028 \; Gev^{3}  \\
\lo P_{1} | \eta \rl = f_{1} m^{2}_{1} \sqrt{ ( m^{2}_{2} + \mu^{2}_{2}
- m^{2}_{ \eta} ) / ( m^{2}_{ \eta'} - m^{2}_{ \eta} ) }
\simeq 0.0019 \; Gev^{3}  \\
\lo P_{1} | \eta' \rl = f_{1} m^{2}_{1} \sqrt{ ( m^{2}_{ \eta'} -
m^{2}_{2} - \mu^{2}_{2} ) / ( m^{2}_{ \eta'} - m^{2}_{ \eta} ) }
\simeq 0.0018 \; Gev^{3}  \\
\lo P_{2} | \eta \rl = - f_{2} m^{2}_{2} \sqrt{ ( m^{2}_{1} + \mu^{2}_{1}
- m^{2}_{ \eta} ) / ( m^{2}_{ \eta'} - m^{2}_{ \eta} ) } \simeq
 0.056 \; Gev^{3}  \\
\lo P_{2} | \eta' \rl = f_{2} m^{2}_{2} \sqrt{ ( m^{2}_{ \eta'}
- m^{2}_{1} -\mu^{2}_{1} ) / ( m^{2}_{ \eta'} - m^{2}_{ \eta})}
\simeq 0.062 \; Gev^{3}
\eea
In the next sections we will see that these are the matrix elements
(30)-(35) that determine all the corrections of interest.
\newpage
\section{ Isospin breaking as a perturbation }
Now we are able to proceed to discussing the isospin breaking effects.
The easiest way to do it is to use a perturbation theory over the
isospin breaking Hamiltonian $ H^{ \Delta I = 1} = \frac{m_{u} - m_{d}}
{2} ( \bar{u} u - \bar{d} d )  $ . Let us write down the total
Hamiltonian as
\bea
H =   H_{QCD} + \frac{ m_{u} + m_{d}}{2} ( \bar{u} u + \bar{d} d )
+ m_{s} \bar{s} s - \theta \frac{ \alpha_{s}}{2 \pi} ( \vec{ \pi}
\vec{H}) + \theta^{2} ( \frac{ \alpha_{s}}{ 2 \pi} )^{2} \vec{H}^{2}
 \nonumber   \\
        + \frac{m_{u} - m_{d}}{2} ( \bar{u} u - \bar{d} d )
\: = \: H^{0} + H^{ \Delta I = 1 }
\eea
where $ \vec{ \pi^{a}} \: = \: \.{\vec{A^{a}}} + \frac{ \alpha_{s}}
{ 2 \pi} \theta \vec{H^{a}} $ is the canonical momentum and $ H_{QCD} $
 corresponds to the massless theory. The $ \theta $ - term displayed in
(36) is just another way of incorporating the Veneziano ghost discussed
above, provided $ \ll Q Q \rl^{W} = -  \partial^{2} \varepsilon_{
vac} / \partial \theta^{2} $  [11,2,3] ( in such formulation, the theory
contains only gauge invariant states ). The Hamiltonian form of the
Veneziano construction enables us to build the perturbation theory
around the $ SU(2)_{V} $ symmetric Witten-Veneziano Hamiltonian $ H^{0} $
. One deals there with the ( gauge invariant ) states $ \pb , \etb ,
\et'b $ defined by (8),(9),(28),(29). Then the first order corrections
to the wave functions over the perturbation $ H^{ \Delta I = 1} $ are
given by the quantum mechanical formula
\beq
| \Psi \rl ' = \sum_{m \neq n } \frac{ V_{mn}}{ E_{n} - E_{m}}
\frac{1}{ \ll \Psi_{m} | \Psi_{m} \rl } | \Psi_{m}  \rl  =
\sum_{ m \neq n} \frac{ V_{mn}}{ 2 E_{m} ( E_{n} - E_{m}) }
| \Psi_{m} \rl
\eeq
where $ V_{mn} $ stands for the matrix element of the perturbation
\beq
V_{mn} = \ll \Psi_{m} ( \vec{p} , E_{m} ) | H^{ \Delta I = 1 } |
\Psi_{n} ( \vec{p} , E_{n} ) \rl
\eeq
The only non-vanishing matrix elements $ V_{ \pi \eta ( \eta') } $
can be evaluated in the pion's rest frame by the soft pion technique
[12] :
\beq
\ll \pi_{B} | H ^{ \Delta I = 1} | \eta_{B} ( \eta'_{B} ) \rl
= \frac{1}{ f_{ \pi}} \frac{ m_{u} - m_{d}}{ m_{u} + m_{d}}
\lo P_{1} | \eta_{B} ( \eta'_{B} ) \rl
\eeq
( Using the soft pion theorem in this case seems to be harmless since
an expected accuracy is $ O( \frac{ m^{2}_{ \pi}}{ ( m_{ \eta} - m_{ \pi}
)^{2} } ) \simeq 10 $ \%  ).
Finally, the wave functions formulae read
\bea
| \pi \rl & = & \pb + \frac{ V_{ \pi \eta}}{ 2 m_{ \eta} ( m_{ \pi}
- m_{ \eta}) } \etb + \frac{ V_{ \pi \eta'}}{ 2 m_{ \eta'} ( m_{ \pi}
- m_{ \eta' } ) } \et'b      \nonumber  \\
|\eta \rl & = & \etb + \frac{ V_{ \pi \eta}}{ 2 m_{ \pi} ( m_{ \eta}
- m_{ \pi} ) } \pb        \\
|\eta' \rl & = & \et'b + \frac{ V_{ \pi \eta'}}{ 2 m_{ \pi} ( m_{ \eta'}
 - m_{ \pi} ) } \pb     \nonumber
\eea
( Of course, all the masses here are to be calculated in the $ SU(2)_{V}
$ limit. Note also that there is no need for changing the normalizations
of the states since these effects are of second order in the perturbation
). Numerically, $ \theta_{ \pi \eta } = \frac{ V_{ \pi \eta}}{2 m_{ \eta}
( m_{ \pi} - m_{ \eta} ) } \simeq 0.032 \ratio $ and the analogous
$ \theta_{ \pi \eta'} \simeq 0.009 \ratio $ . It should be stressed that
these mixing angles are quite different from those discussed usually for
the mixing among the {\it massless} states [1]. Moreover, any calculation
scheme based on treating the total mass term $ L^{m} = - m_{u} \bar{u}u
- m_{d} \bar{d} d $ as a perturbation seems to be in troubles when
concerns isospin breaking matrix elements. The same mass term is
responsible there both for the mixing and for non-vanishing values
of matrix elements for "bare" states, thus the problem of double
counting arises. In contrast, there is no such difficulty in the
present approach. Let us point out also that as long as all the
diagonal matrix elements $ V_{nn} $ vanish, the mass corrections
start only with the second order and, thus, are very small. For
example, the contribution of this mechanism into $ \pi^{+} -
\pi^{0} $ mass difference is only about 0.18 $ Mev $ for $ \ratio
\simeq 0.4 $ .

Now we can, using (30),(31),(39),(40) , calculate the pion matrix
element of the topological density :
\beq
\mgluon = 2 ( \theta_{ \pi \eta} \lo Q \etb + \theta_{ \pi \eta'}
 \lo Q \et'b )
\eeq
This formula can be easily cast into a form comparable with (2).
One obtains the following formula for the correction factor $ \zeta $
\beq
\zeta = \frac{ \mu_{1} ( \mu_{1} + \mu_{2})}{2 ( \mu^{2}_{1} + \mu^{2}_{
2} ) } -1 + O ( m^{2}_{1}, m^{2}_{2} ) \simeq 0.6 - 1 + 0.003 \simeq
-0.4
\eeq
( the number 0.003 is the contribution of the $ O ( m^{2}_{s} ) $ term
omitted in (42) ) . We see that the naive answer (2) widely used
in the literature [5],[6] is reduced by about 40 \% . The couplings
of the ghost-goldstone interactions $ \mu_{1}, \mu_{2} $ constitute
the hidden large parameters in the problem. Note that $ \eta $ and
$ \eta'$ contribute about equally into (41).

\section{ Strange quark content of pion}

This is a good exercise to get a feeling of a size of effects that
the above considered mechanism can bring in. Let us start with
the one-pion matrix elements. In this case, non-vanishing values
for them are obtained as a sole result of the mixing. One finds
from (34),(35),(40)
\beq
\lo P_{2} \rpi \simeq  \frac{1}{2} \ratio m^{2}_{ \pi} ( -0.176
+ 0.058 )  \simeq - \ratio m^{2}_{ \pi}  \cdot 0.06
\eeq
where the two numbers in the parenthesis stand for the numerical
contributions of the $ \eta $ and $ \eta'$ , correspondly. Let
us point out that the value (43) is again about twice less than
the original estimate of GTW [1]. The axial current matrix
element $ \lo \bar{s} \gmmu \gmf s \rpi = i f_{ \pi} A q_{ \mu} $
can be easily found from (41), (43) :
\beq
A = \frac{1}{ f_{ \pi} m^{2}_{ \pi} } \lo P_{2} + 2 Q \rpi
\simeq - \ratio  \cdot 0.03
\eeq
One may further address the strange quark condensate in the pion.
One obtains
\bea
\ll \pi | \bar{s} s | \pi \rl &=& \ll \pi_{B} | \bar{s} s | \pi_{B}
\rl + \theta^{2}_{ \pi \eta} \ll \eta_{B} | \bar{s} s | \eta_{B} \rl
+ \theta^{2}_{ \pi \eta'} \ll \eta'_{B} | \bar{s} s | \eta'_{B} \rl
 \nonumber  \\
                             &  & + 2 \theta_{ \pi \eta } \theta_{
 \pi \eta'} \ll \eta_{B} | \bar{s} s | \eta' \rl
\eea
Let us first concentrate on the correction terms in (45). To find
$ \ll \eta_{B} | \bar{s} s | \eta_{B} \rl \: , \: \ll \eta_{B} |
\bar{s} s | \eta'_{B} \rl $ , one can apply the "soft $ \eta$ -
theorem" ( we neglect here the mixing with the singlet ) :
\bea
\ll \eta_{B} | \bar{s} s | \eta_{B} \rl = - \frac{2}{ \sqrt{6}
f_{ \eta} m_{s}} \lo P_{2} | \eta_{B} \rl \simeq 2.1 \; Gev  \\
\ll \eta_{B} | \bar{s} s | \eta'_{B} \rl = - \frac{2}{ \sqrt{6}
f_{ \eta} m_{s}} \lo P_{2} | \eta'_{B} \rl \simeq 2.3 \; Gev
\eea
One can argue ( by comparing (46),(47) with an analog of (49) for
$ \eta \leftrightarrow \eta' $ ) that the accuracy of these
estimates is of order 30 \% . The calculation of $ \ll \eta'_{B}
| \bar{s} s | \eta'_{B} \rl $ is slightly more involved. Consider
the correlation function
\beq
\Pi = i \int dx \lo T \{ P_{2} (x) \bar{s} s (0) \} | \eta'_{B}
\rl
\eeq
The pole approximation applied to (48) can be shown to give the
following relation
\beq
\frac{1}{m^{2}_{ \eta'}} \lo P_{2} + 2 Q | \eta'_{B} \rl
\ll \eta'_{B} | \bar{s} s | \eta'_{B} \rl + \frac{1}{m^{2}_{ \eta}}
\lo P_{2} + 2 Q | \eta_{B} \rl \ll \eta_{B} | \bar{s} s | \eta'_{B}
\rl
= \frac{1}{m_{s}} \lo P_{2} | \eta_{B} \rl
\eeq
Together with (46),(47) this yields
\beq
\ll \eta'_{B} | \bar{s} s | \eta'_{B} \rl \simeq 1.2 \; Gev
\eeq
Thus, the mixing contribution into (45) is estimated as
\beq
\frac{ \ll \bar{s} s \rl_{ \pi}}{ \ll \bar{u} u  + \bar{d} d
\rl_{ \pi}}|_{ ( mixing)} \simeq 3.3  \cdot 10^{-5}
\eeq
This estimate is to be compared with the first term in (45). It
has been recently calculated ( for the charged pions) within
the chiral perturbation theory [13] and the NJL model [14] :
\beq
\frac{ | \ll \bar{s} s \rl_{ \pi} | }{ \ll \bar{u}u + \bar{d}d
 \rl_{ \pi} } \leq ( 4 - 6 ) \cdot 10^{-4}
\eeq
that apparently is of one order larger than the mixing
contribution (51). The unpleasant thing with (52) is that
this ratio turns out very sensitive to the choice of
$ \hat{m} = 1/2 ( m_{u} + m_{d} ) $ and even changes the
sign when $ \hat{m} $ varies from 5.1 to 5.8 Mev [14].
So, one may wonder whether this value is not exactly zero, up
to an accuracy of the methods under consideration. One can
argue that this is not the case from the following argument.
Let us differentiate the WI (10) ( taken in the $ SU(2)_{V} $
 limit ) over the mass of the strange quark :
\beq
- i \ll P_{3} \: \bar{s} s \: P_{3} \rl = - 4 m_{q} \frac{ d \ll
\bar{q} q \rl }{ d m_{s}}
\eeq
The dominant contribution into the l.h.s. of (53) is due to the
pion intermediate state, so that we arrive at
\beq
\lpi \bar{s}s \rpi = \frac{4 m_{q}}{ f^{2}_{ \pi}} \frac{ d \ll
\bar{q} q \rl }{ d m_{s}}
\eeq
The quantity $ d \ll \bar{q} q \rl / d m_{s} $ ( $ \ll \bar{q} q
\rl = \ll \bar{u}u \rl $ ) in the chiral SU(2) limit has been
calculated in Ref.[19] within the instanton vacuum model:
$ K = d \ll \bar{u} u \rl / d m_{s} \simeq - 0.085 \: Gev^{2} $ .
Assuming that the correlator K in the $ SU(2)_{V} $ limit is not
very different from this value, one obtains from (54)
\beq
\frac{ \ll \bar{s} s \rl _{ \pi} }{ \ll \bar{u} u + \bar{d} d \rl
_{ \pi}} \simeq - 3 \cdot 10^{-2}
\eeq
We conclude that the mixing only slightly affects the strange quark
condensate in the pion. Still, a large discrepancy between two
estimates (52) and (55) deserves further studying.

\section{ Isospin breaking in the $ \pi N $ interaction }

Another interesting consequences of the above developed formalism
concern isospin symmetry breaking effects in the $ \pi N $ and NN
interactions, the long staying problem in nuclear physics. Among
them, one frequently discussed phenomenon is the scattering length
difference in the $ ^{1}S_{0} $ partial wave ( $ |a_{nn} | -
| a_{pp} | = 1.5 \pm 0.5 \: fm $  with the Coulomb corrections
subtracted ). Another one is the discrepancy of order a few hundreds
keV between the measured masses of the mirror nuclea after allowing
for the n-p mass difference and the calculated e.m. corrections -
this is the so-called Nolen-Schiffer (NS) anomaly [15] ( see [16,23]
for review ). The data signals that the nn interaction is more
attractive than the pp one , that in turn means $ | g_{ \pi_{0} nn} |
> g_{ \pi_{0} pp } $ in the framework of the one-pion exchange
potential models ( OPEP ) . However, the attempts of direct evaluating
the coupling $ g_{ \pi_{0} pp} $ from a phase shift analysis suffer
large uncertainties [16], so that no decisive conclusion can be driven.
The situation in the theory is also ambiguous since the various quark
model based calculations disagree even in the signs [16].

 Let us now appeal to our scheme. For the Yukawa Lagrangian
\beq
L_{Yuk}= i g^{(0)}_{ \pi} ( \bar{p} \gmf p - \bar{n} \gmf n )
\pi^{0}_{B} + i g^{(0)}_{ \eta} ( \bar{p} \gmf p + \bar{n} \gmf
n ) \eta_{B} + ( \eta_{B} \leftrightarrow \eta'_{B} )
\eeq
( the $ SU(2)_{V} $ limit is implied ) one obtains from (40)
\beq
|g_{ \pi^{0}pp ( \pi^{0} nn ) } | = g^{(0)}_{ \pi} \pm
( \theta_{ \pi \eta} g^{(0)}_{ \eta} + \theta_{ \pi \eta'}
g^{(0)}_{ \eta'} )
\eeq
As the both mixing angles are positive, it is seen from (57) that the
mixing tends to increase the coupling $ g_{ \pi^{0}pp} $ and reduce
$ g_{ \pi^{0} nn} $ , i.e. yields a result {\it opposite} to what is
expected for explaining the isospin asymmetry. Unfortunately, it is
not so easy to determine exactly the $ SU(2)_{V} $ limit of the
coupling constants $ g_{ \eta} , g_{ \eta'} $ , though some estimates
can still be done.

 The first one is the SU(3) prediction $ g_{8} = \sqrt{3} ( \: 1 -
\frac{4}{3}
D/(D+F) \: ) g_{ \pi} $ . Substituting $ D/( D+F) \simeq 0.6 $ from the
hyperon decay data [17] and neglecting the octet-singlet mixing, one
finds $ g_{ \eta} \simeq g_{8} \simeq 4.6 $ . The second estimate comes
from the Goldberger-Treiman relation for the $ \eta $ :
\beq
2 M ( \Delta u + \Delta d - 2 \Delta s ) = \sqrt{6} f_{ \eta} g_{ \eta}
\eeq
With $ \Delta u + \Delta d - 2 \Delta s ) \simeq 0.68 $  [17] and
$ f_{ \eta} \simeq 0.6 f_{ \pi} $ from th $ \eta \rightarrow 2
\gamma $ decay, this yields $ g_{ \eta} \simeq 6.3 $ .

One more estimate for the couplings $ g_{ \eta} , g_{ \eta'} $ can
be obtained under assumption on their slight variation with moving
from the SU(2) to SU(3) chiral limit. Let us apply the pole
approximation to the matrix element of the topological density over
the proton states $ \ll p | Q | p \rl = \lim_{ p' \rightarrow p }
\ll p' | Q | p \rl $ . Then
\beq
\ll p | Q | p \rl = - ( \frac{ \lo Q | \eta \rl }{ m^{2}_{ \eta} }
g_{ \eta} + \frac{ \lo Q | \eta' \rl }{ m^{2}_{ \eta'} } g_{ \eta'}
) \bar{p} i \gmf p
\eeq
On the other hand, this quantity has been estimated in Ref.[18] by
some extension of an argumentation based on the dimensional
transmutation phenomenon :
\beq
\ll p | Q | p \rl = - \frac{ 2 n_{f} }{ 3 b } m_{p} \bar{p} i \gmf p
\; , \; b = \frac{11}{3} N_{c} - \frac{2}{3} n_{f}
\eeq
( $ n_{f} $ stands for the number of massless flavors ). Assuming the
validity of this formula, one can obtain
\bea
0.032 g_{ \eta} + 0.031 g_{ \eta'} = 0.123 \; ( \: chiral \: SU(2) \:) \\
0.027 g_{ \eta} + 0.038 g_{ \eta'} = 0.102 \; ( \: chiral \: SU(3) \:) \:
,
\eea
where we have also accounted for the fact that the s-quark carry about
one half of the nucleon mass [19,20]. ( Note that the first term in the
r.h.s. of eq. (59) survives the chiral SU(3) limit.) Then one obtains
the estimate
\beq
g_{ \eta} \simeq 3.9 \; , \; g_{ \eta'} \simeq 0
\eeq
( Taken literally, the system (61),(62) yields $ g_{ \eta} = 4.01 \: ,\:
g_{ \eta'} = -0.17 $ .) It is worth noting that estimate (63) is
consistent with the claims on a smallness of the $ g_{ \eta'} $ coupling
[21,22] driven from the spin crisis studies. The value $ g_{ \eta} \simeq
3.9 $ appears to be a lower bound for the $ SU(2)_{V} $ value $ g^{(0)}_{
\eta} $ . Thus, our final estimates are
\beq
g^{(0)}_{ \eta} \simeq 5 \pm 1 \; , \; g^{(0)}_{ \eta'} \simeq 0
\eeq
{}From (57),(64) we find ( $ \frac{ g^{2}_{ \pi N N } }{ 4 \pi}
\simeq 14 $ )
\bea
\frac{ g_{ \pi^{0} pp} - | g_{ \pi^{0} nn }| }{ g_{ \pi NN }}
\simeq \ratio ( 2.4 \pm 0.5 ) \cdot 10^{ -2} \; , \;     \\
- \alpha  \equiv \frac{ g^{2}_{ \pi^{0} pp} - g^{2}_{ \pi^{0} nn}}{
g^{2}_{ \pi NN} } \simeq ( 1.9 \pm 0.4 ) \cdot 10^{ -2} \; , \;
\; \ratio \simeq 0.4   \nonumber
\eea
Thus, the long-range part of the
          charge asymmetric nuclear potential $ V_{CA} = V_{nn} -
V_{pp} $ for the $ ^{1}S_{0} $ state can be written as
\beq
V_{CA} = \frac{ g^{2}_{ \pi^{0}pp} - g^{2}_{ \pi^{0}nn}}{ g^{2}_{ \pi
NN}} \frac{ g^{2}_{ \pi NN }}{ 4 \pi} \frac{ m^{2}_{ \pi}}{ 2 M^{2}_{N}}
\frac{ e^{ - m_{ \pi} r }}{r}
\eeq
It is known that for explaining the NS anomaly for A = 41 , $ \alpha $
must be larger than 2 \% [23]. The binding energy difference for A = 41
calculated within the shell model wave functions provides $ E_{ ^{41}Sc}
- E_{ ^{41}Ca} \simeq - 780 \: keV $ that is the right number in modulus
but wrong in the sign ! ( The answer for the shell model matrix element
was kindly presented to us by N.Auerbach .)

A few comments are in order here. From the viewpoint of the $ 1/N_{c} $
expansion, the presented effect is $ O ( N_{c} ) $ .
Thus, it has to be taken on equal footing with other $ O ( N_{c} ) $
contributions which can be found e.g. from the the chiral theory of
the nucleon or the QCD sum rules. Unfortunately, this can be done
within the modern state-of-art only for the strangeless nucleon,
whereas the strangeness is effectively appearing in our calculations.
We believe, however, that our estimate (66) constitutes the leading
contribution into the charge asymmetric nuclear potential as resulting
from a strong interaction of the ghost.
Another observation is that our
conclusion on the sign of ( $ g_{ \pi^{0} pp} - g_{ \pi^{0} nn } $ )
is not at variance with most quark model based calculations ( the only
exception is the cloudy bag model ) [16]. In contrast to previous works,
it is deduced this time from the exact QCD dynamics. In all likelihood,
the result (65),(66) means that long-range meson exchange forces cannot
explain the isospin violation in the $ \pi N $ and NN interactions. On
the other hand, the large effect of (66) has to be taken into account
in any
scheme of resolving the NS anomaly, e.g. the one based on a partial
restoration of the chiral symmetry in the nuclear medium (see e.g. [24]).

\section{ Proton spin and violation of Bjorken sum rule }

Now we would like to come back to the above mentioned connection of the
discussed phenomena with the spin crisis problem ( see e.g. [25] for
review ) . The necessity of explaining the data on deep inelastic
polarized lepton-nucleon scattering has led to re-examining some
usually done assumptions such as e.g. an isoscalarity of the
anti-quark sea in the nucleon. In this view, there has been raising
interest [26] during a last few years in checking in QCD the Bjorken
sum rule [27] , relating the isotriplet component of the first moment
of the polarized nucleon structure function $ g_{1} (x) $ to the weak
coupling $ g_{A} $ . The validity of the Bjorken sum rule has been
discussed in the recent papers [28,29]. As it is argued in Ref. [28],
the Bjorken sum rule is consistent with all the available data at the 12
\% level after the kinematic and higher-twist power corrections [30] are
taken into account. These corrections have been however calculated in the
chiral SU(2) limit, thus leaving open the question on a possible
violation of the Bjorken sum rule due to isospin breaking. This latter
point has been discussed in Ref.[29] under assumption on the validity
of the Sutherland theorem in the isoscalar channel [1] which is to be
abandoned according to the results of the present work. Thus, it would be
interesting to estimate corrections to the Bjorken sum rule resulting
from above considered isospin breaking in the meson couplings.

The operator product expansion of the antisymmetric part of the
T-product of two electromagnetic currents provides
\beq
\int dx g^{N}_{1} ( x , Q^{2} ) = \frac{1}{2} ( 1- \frac{ \alpha_{s}
( Q^{2}) }{ \pi} + O( \alpha^{2}_{s}) ) [ \frac{4}{9} \Delta u^{N}
+ \frac{1}{9} \Delta d^{N}
 + \frac{1}{9} \Delta s^{N} ] ( Q^{2} ) + O ( \frac{ M^{2}}{Q^{2}} )
\eeq
where
\beq
\ll N | ( \bar{q}_{i} \gmmu \gmf q_{i} )^{ \mu^{2} = Q^{2} }
| N \rl  =  \Delta q_{i} ( Q^{2}) s_{ \mu}
\eeq
and $ s_{ \mu} = \bar{u} (p,s) \gmmu \gmf u( p,s) $ is the proton
spin vector. Let us estimate first the s-quark contribution into
the nucleon spin. To this end, consider the matrix element
\beq
\ll N(p') | \bar{s} \gmmu \gmf s | N(p) \rl =
\bar{u} (p') [ \gmmu \gmf G^{N}_{1} ( q^{2} ) + q_{ \mu}
\gmf G^{N}_{2} ( q^{2} ) ] u(p)
\eeq
where $ q = p' - p $ . Differentiating this relation and saturating
it with the  $ \pi , \eta , \eta' $ contributions, we find in the
limit $ q \rightarrow 0 $
\beq
\Delta s^{N} = G^{N}_{1} ( 0) = \frac{1}{ 2 m_{N} } \sum_{ i=
\pi , \eta, \eta' } \frac{ \lo P_{2} + 2  Q | i \rl }{ m^{2}_{i}}
g_{i NN}
\eeq
Then, using (30),(31),(34),(35),(40),(44),(63), we obtain
\beq
\Delta s^{p} \simeq - 0.30 \; , \; \Delta s^{n} \simeq -0.28
\eeq
( These values refer to a low normalization point $ \mu \sim 500
\: Mev $ . We neglect, however, a weak logarithmical dependence
on $ Q^{2} $ due to the anomalous dimensions. ) Note that the
dominant contribution into (69) comes from the $ \eta $ with
$ g_{ \eta nn ( \eta pp ) }  \simeq g^{(0)} \pm 0.66 $ ( see (40) ).
One can conclude from (71) that the s-quark fraction of the nucleon
spin is close to the isoscalar and thus cannot bring any sizable
correction into the Bjorken sum rule. Numerically, the answer for
$ \Delta s^{p} $ is somewhat larger than the latest value
$ \Delta s^{p} = -0.13 \pm 0.04 $ [28] or the previous world
average $ \Delta s^{p} = -0.20 \pm 0.11 $ [31] , which have been
obtained within the SU(3) limit. One could expect an accuracy
of order 30 \%  for the estimates (70), which is typical in
applications of the PCAC technique for the kaons. We believe,
however, that the actual accuracy of (70) is in fact better
since it is the approximation that has been used systematically
in the previous discussion of the Veneziano mechanism.

The fraction of the nucleon spin due to the non-strange quarks
can be calculated analogously. It is convenient to decompose
the quark part of the operator in the matrix elements into
the isovector and isoscalar components ( note that we do not
separate parton and anomalous contributions into the
nucleon spin ) :
\beq
m_{u} \bar{u}i \gmf u + m_{d} \bar{d} i \gmf d =
\frac{m_{u} - m_{d}}{2} ( \bar{u}i \gmf u - \bar{d} i \gmf d )
+\frac{ m_{u} + m_{d}}{2} ( \bar{u} i \gmf u + \bar{d} i \gmf d )
\eeq
Then we observe the strong cancellation between different
contributions into $ \Delta u^{p} + \Delta d^{p} $ :
\beq
\Delta u^{p} + \Delta d^{p} \simeq \frac{1}{ m_{p}} ( -0.50
+0.40 + 0.31 - 0.02 ) \simeq 0.21 \: ,
\eeq
where the first two terms in the parenthesis are originating from the
isovector and isoscalar pion matrix elements, the third and fourth
ones are due to the $ \eta $ and $ \eta' $ , correspondly. All the
other contributions are small. Similarly, one obtains for the neutron
\beq
\Delta u^{n} + \Delta d^{n} \simeq \frac{1}{m_{n}}( 0.49 - 0.41
+0.41 + 0.02 ) \simeq 0.54
\eeq
These numbers are to be compared with the ones obtained from the
 SU(3) symmetric fit under assumption on the validity of the
Bjorken sum rule [28]
\beq
\Delta u = 0.80 \pm 0.04 \; , \; \Delta d = - 0.46 \pm 0.04
\eeq
Now we are able to estimate the isospin violating corrections
to the Bjorken sum rule. Neglecting negligible ( of order $ 10^{-4} $
 ) isospin breaking in the isovector part of the e/m current and
the strange current, we obtain after using the Goldberger-Treiman
relation
\bea
\int dx ( g^{p}_{1} (x) - g^{n}_{1} (x) ) \simeq \frac{1}{2}
( 1 - \frac{ \alpha_{s}}{ \pi} )  [ \: \frac{1}{3} g_{A} +
\frac{5}{18} ( \Delta u^{p-n} + \Delta d^{p-n} ) \: ]  \nonumber  \\
 \simeq \frac{1}{2} ( 1 - \frac{ \alpha_{s} }{ \pi} ) [ \frac{1}{3}
g_{A} - 0.09 ]
\eea
Thus, one can expect isospin breaking of order 20 \% in the Bjorken sum
rule. Actually, the formula (76) represents the isospin violation due
to the one-meson reducible contributions [29] into the matrix
element (68). An interesting possibility is that irreducible (e.g.
instanton-induced ) contributions turn out also large [29] but of
opposite sign, thus reducing the total isospin violation in the
Bjorken sum rule. Presumably, the role of the non-resonant contribution
into the proton spin could be clarified in the framework of the
dispersion approach.
\newpage
\section{Conclusions}

Let us summarize the results of this work. We have calculated the mixing
in the $ \pi^{0} - \eta - \eta' $ system basing on the Veneziano
solution of the U(1) problem in the $ SU(2)_{V} $ symmetric world.
The ghost interacts strongly with the pseudo-goldstone states ( or,
more precisely, the  OZI goldstone modes, in the terminology of Ref.[22])
and must be taken into account before allowing for a isospin violation.
As a consequence  of this scheme, we have found large corrections to
the naive estimations of the isospin breaking matrix elements.
In all the cases, the ghost suppresses the isospin violation due to the
quark mass difference. Being applied to the charge asymmetry
phenomena in nuclear physics, our results indicate essential troubles
in attempts of explaining them in terms of meson exchange forces.
For the spin crisis problem, the presented mechanism allows one to
re-estimate "phenomenologically" ( i.e. in the pole approximation ) the
quark contribution into the
proton spin. At the 30 \% level, the obtained values agree with those
derived from the SU(3) symmetric fit. It would be interesting to find
a correspondence between the present work and the effective chiral
Lagrangian technique. We are planning to return to this problem
elsewhere.

The author is grateful to N.Auerbach, M.Karliner and U.Maor for useful
discussions and interest to this work. The special thanks are due to
L.Franktfurt, without whose patience and stimulating discussions
this paper would never been written.
%
%
\newpage
\bibliographystyle{plain}

\end{document}